\documentclass[twocolumn,amssymb,amsmath,floats,showpacs,superscriptaddress,pre]{revtex4}

\usepackage{amsmath,amssymb,bm}
\usepackage{graphics}
\usepackage{epsfig}
\usepackage{graphicx}

\begin{document}

\title{Zero Pearson Coefficient for Strongly Correlated Growing Trees}
\author{S. N. Dorogovtsev}
\affiliation{Departamento de F{\'\i}sica, I3N, Universidade de Aveiro, 3810-193 Aveiro,
Portugal}
\affiliation{A. F. Ioffe Physico-Technical Institute, 194021 St. Petersburg, Russia}
\author{A. L. Ferreira}
\affiliation{Departamento de F{\'\i}sica, I3N, Universidade de Aveiro, 3810-193 Aveiro,
Portugal}
\author{A. V. Goltsev}
\affiliation{Departamento de F{\'\i}sica, I3N, Universidade de Aveiro, 3810-193 Aveiro,
Portugal}
\affiliation{A. F. Ioffe Physico-Technical Institute, 194021 St. Petersburg, Russia}
\author{J. F. F. Mendes}
\affiliation{Departamento de F{\'\i}sica, I3N, Universidade de Aveiro, 3810-193 Aveiro,
Portugal}

\begin{abstract}

We obtained Pearson's coefficient of strongly correlated recursive networks growing by preferential attachment of 
every new vertex by $m$ edges. We found that the Pearson coefficient is exactly zero in the infinite network limit for the recursive trees ($m=1$). If the number of connections of new vertices exceeds one ($m>1$), then the Pearson coefficient in the infinite networks equals zero only when the degree distribution exponent $\gamma$ does not exceed $4$. 
We calculated the Pearson coefficient for finite networks and observed a 
slow, 
power-law like approach to an infinite network limit. Our findings indicate that Pearson's coefficient 
strongly depends on size and details of networks,  
which makes this characteristic virtually useless for quantitative comparison of 
different 
networks. 

\end{abstract}

\maketitle





\begin{figure}[b]
\begin{center}
\scalebox{1.123}{\includegraphics[angle=0]{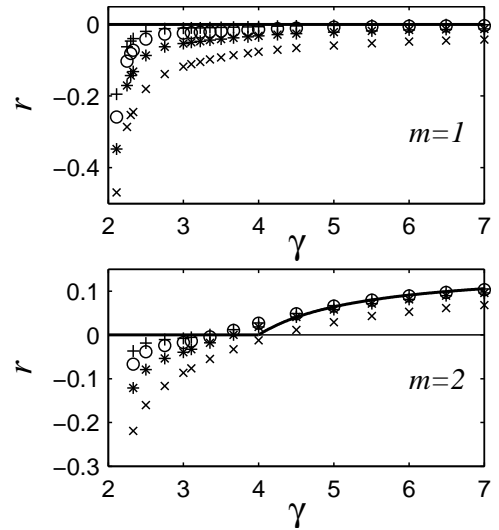}}
\end{center}
\caption{ 
Pearson's coefficient $r$ vs. degree distribution exponent in the recursive networks with $m=1$ and $2$ for the networks of $10^3$ ($\times$), $10^4$ ($\times{}\hspace{-7.25pt}+$), $10^5$ ($\circ$), and $10^6$ ($+$) vertices. 
The thick lines show the Pearson coefficient for the infinite networks. 
} 
\label{f1}
\end{figure}

The Pearson coefficient $r$ is used as an integral characteristic of structural correlations in a network. Pearson's coefficient characterizes pairwise correlations between degrees of the nearest neighboring vertices in networks. Some observable quantities in correlated networks (e.g., the size of a giant connected component near the point of its emergence) are directly expressed in terms of this coefficient \cite{Dorogovtsev:dgm08,Goltsev:gdm08}. The Pearson coefficient is the normalized correlation function of the degrees of the nearest neighbour vertices 
\cite{Newman:n02b,Newman:n03,Serrano:sbpv07}. The coefficient is the ratio: 
\begin{equation}
r = 
\frac{\langle jk \rangle_{\text{e}}-\langle k \rangle_{\text{e}}^2}
{\langle k^2 \rangle_{\text{e}}-\langle k \rangle_{\text{e}}^2}
= 
\frac{\langle(j - \langle k \rangle_{\text{e}})(k - \langle k \rangle_{\text{e}})\rangle_{\text{e}}}{\sigma^2}
, 
\label{e10}
\end{equation}
where $\langle jk \rangle_{\text{e}}$ is the 
average product of the degrees $j$ and $k$ of the end vertices of an edge, 
$\langle k \rangle_{\text{e}}=\langle k^2 \rangle/\langle k \rangle$ is 
the average degree of an end vertex of an edge, 
$\langle k^2 \rangle_{\text{e}}=\langle k^3 \rangle/\langle k \rangle$ is 
the average square of the degree of an end vertex of an edge \cite{bookmark1}, 
and 
\begin{equation}
\sigma^2 \equiv \frac{\langle k^3 \rangle}{\langle k \rangle} - 
\frac{\langle k^2 \rangle^2}{\langle k \rangle^2}
\label{e20}
\end{equation}
is for normalization. 
Here $\langle \ldots \rangle$ and $\langle \ldots \rangle_{\text{e}}$ denote averaging over vertices and edges, respectively [see Eqs.~(\ref{e150}), (\ref{e170}), and (\ref{e185}) below]. 
Pearson's coefficient can be positive (in average, assortative mixing of the nearest neighbors' degrees) or negative (disassortative mixing) and takes values in the range from $-1$ to $1$. 
The Pearson coefficient $r$ is a convolution of the joint distribution of nearest neighbor degrees, $P(j,k) \equiv e_{jk}$. 
This joint distribution is the probability that the ends of a randomly chosen edge have degrees $j$ and $k$, $\sum_{jk}e_{jk}=1$.  
Being an integral characteristic of degree--degree correlations, the Pearson coefficient misses details of these correlations, compared to $e_{jk}$ \cite{Pastor-Satorras:pvv01,Vazquez:vpv02,Maslov:ms02}. 
Despite this fact, Pearson's coefficient is widely used for characterization and comparison of real-world networks \cite{Newman:n03}. 
Note that the compared real networks have different sizes. 
In this paper we show that since $r$ is a markedly size dependent quantity, Pearson's coefficient may be used for comparison of networks only with a very critical attitude.  
We calculate Pearson's coefficient for the simplest growing 
complex networks, namely recursive random networks with preferential attachment of new vertices. 
We 
describe 
the size dependence of $r$ in 
these networks, Fig.~\ref{f1}. 
Remarkably, for all infinite recursive trees of this kind, we find that the Pearson coefficient is exactly zero 
at any value of the degree distribution exponent $\gamma$,  
although all these networks are strongly correlated. 
The statement that Pearson's coefficient is zero in the range $\gamma>4$ is essentially non-trivial. 
The point is that zero value of the Pearson coefficient in the range $\gamma\leq 4$ where the third moment of the degree distribution diverges is clear. Indeed, in this region, the denominator in definition (\ref{e10}) diverges in the infinite network limit \cite{bookmark3}. 
This is not the case at $\gamma>4$, and for zero $r$, the numerator in Eq.~(\ref{e10}) must be zero. In this range, zero value of Pearson's coefficient means a surprising exact mutual compensation of different degree--degree correlations in these trees.  
There are two opposing kinds of degree--degree correlations in complex networks: assortative and disassortative. Here assortativity is a tendency of high degree vertices to have high degree neighbors and low-degree vertices to have low degree neighbors. 
In contrast, disassortative mixing means neighborhood of vertices with contrasting (low and high) degrees. This tendencies may be opposing in different ranges of degrees, see discussion in Sec.~\ref{zz}, Fig.~\ref{f4}. That is, assortative and disassortative mixing may coexist. In particular, this is the case for random recursive trees. Our results show that in these growing networks, the different kinds of correlations completely compensate each other in the infinite network limit.

\section{Main results} \label{results}

We study recursive networks in which 
each new vertex is attached to $m$ existing ones chosen with probability proportional to a linear function of vertex degree, $k+A \equiv k+am$. This rule generates scale-free correlated networks with a degree distribution exponent $\gamma=3+A/m=3+a$. 

In the infinite network limit, Pearson's coefficient $r_\infty(\gamma\leq 4)=0$, since the third moment of the degree distribution [see the denominator of Pearson's coefficient definition (\ref{e10})] diverges. 
Here we define $r_\infty \equiv r(t{\to}\infty)$, where $t$ in the number of vertices in a network. For $\gamma>4$, we find 
\begin{equation}
r_\infty = 
\frac{(m{-}1)(a{-}1)[2(1{+}m) + a(1{+}3m)]}{(1{+}m)[2(1{+}m) + a(5{+}7m) + a^2(1{+}7m)]}
, 
\label{e30}
\end{equation}
which shows that for $m=1$ (i.e., for random recursive trees), Pearson's coefficient $r_\infty=0$ for any value of $\gamma$. One can also see that $r_\infty(\gamma=4)=0$ for any $m$. 
In particular, for uniformly random attachment, i.e., $\gamma{\to}\infty$, we have: 
\begin{equation}
r_\infty(\gamma{\to}\infty) = \frac{(m - 1)(1 + 3 m)}{(1 + m)(1 + 7 m)}
.  
\label{e40}
\end{equation}

\begin{figure}[t]
\begin{center}
\scalebox{0.24}{\includegraphics[angle=0]{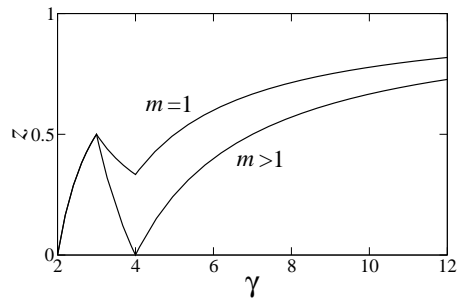}}
\end{center}
\caption{ 
Variation of exponent $z$ [$\delta r(t) = r(t)-r_\infty \propto t^{-z}$] with $\gamma$ exponent of a degree distribution. 
} 
\label{f2}
\end{figure}

\begin{figure}
\begin{center}
\scalebox{0.84}{\includegraphics[angle=0]{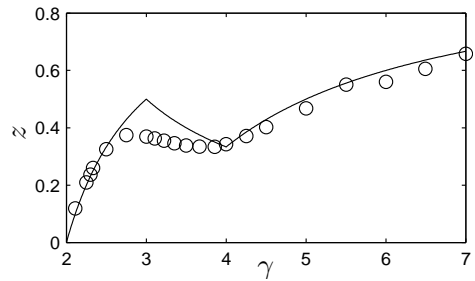}}
\end{center}
\caption{ 
The measured dependences of exponent $z$ on $\gamma$ for recursive trees ($m{=}1$). 
The solid curve is the theoretical dependence $z(\gamma,m{=}1)$. 
The data points are 
obtained 
by power-law fitting 
of the
Pearson coefficient of simulated trees with sizes close to the largest
simulated size ($10^6$ vertices). 
} 
\label{f3}
\end{figure}

Figure~\ref{f1}, obtained by simulation of the networks of $10^3$, $10^4$, $10^5$, and $10^6$ vertices 
(the number of runs for each point was between 50 and 500 for $\gamma<3$ and
between $5\times 10^3$  and $10^4$ for $\gamma>3$), demonstrates the size dependence of Pearson's coefficient in these networks. This figure shows that even for large networks,  
the deviations from the limiting infinite network values are significant. 
We find the following asymptotic size dependences $\delta r(t)=r(t)-r_\infty$: 
\begin{equation}
\delta r(t) \propto \left\{ 
\begin{array}{l}
-
t^{-(\gamma-2)/(\gamma-1)}, \ \text{any } m, \ \ \ \ \gamma<3
, 
\\[10pt]
\left.
\begin{array}{l}
\!\!-
t^{-1/(\gamma-1)}, \ \ \ \ \ \ m=1
,
\\[5pt]
\!\!+
t^{-(4-\gamma)/(\gamma-1)}, \ m>1 
\end{array}
\right\} 
\ 3<\gamma<4 
,
\\[21pt]
\left.
\begin{array}{l}
\!\!-t^{-(\gamma-3)/(\gamma-1)}, \ m=1, 
\\[5pt]
\!\!+t^{-(\gamma-4)/(\gamma-1)}, \ m>1 
\end{array}
\right\} 
\ \gamma>4
.
\end{array}
\right.
\label{e50}
\end{equation} 
Here we show 
the signs of the asymptotes but ignore their factors. 
The positive sign of the
asymptotes at $m{>}1$, $3{<}\gamma{<}4$ and $\gamma{>}4$ means that $r(t)$ approaches the infinite size limit $r_\infty$ from above. 
In this situation ($m{>}1$, $\gamma{>}3$), Pearson's 
coefficient varies with size non-monotonously: first increases and then diminishes to $r_\infty$ (see the second panel of Fig.~\ref{f1}). 
Introducing exponent $z$: $\delta r(t) \propto t^{-z}$ for large network sizes $t$, we arrive at the dependences $z(\gamma)$ shown in Fig.~\ref{f2}. Note that $z<1$, so the infinite network limit is approached slowly. The relaxation to the infinite network values, $r_\infty$, is especially slow if exponent $\gamma$ is close to $2$ (at any $m$) or to $4$ (only if $m>1$). In the specific case of $m>1$, $\gamma=4$, we obtain the logarithmic relaxation: 
\begin{equation}
\delta r(t) \propto \frac{1}{\ln t}
.  
\label{e60}
\end{equation}
We measured the dependence of exponent $z$ on $\gamma$ 
in the simulated networks. The result, shown in Fig.~\ref{f3}, demonstrates an agreement with above analytical predictions. 

\section{Pearson coefficient of infinite networks} \label{infinite}

Let us derive the Pearson coefficient (\ref{e50}) of the infinite recursive networks. The derivation is based on rate equations for the average number $N_k(t)$ of vertices of degree $k$ in a recursive network at time $t$ and the average number $E_{jk}(t)$ of edges connecting vertices of degrees $j$ and $k$. We have   
\begin{equation}
\sum_{k=m} N_k(t) = t
\label{e70}
\end{equation}
for vertices, and so
\begin{equation}
N_k(t) = tp_k(t)
, 
\label{e80}
\end{equation}
where $p_k(t) \equiv P(k,t)$ is a degree distribution of the network of size $t$. 

For edges, we have: 
\begin{equation}
\frac{1}{2}\sum_{j,k}E_{jk}(t) \cong mt
,
\label{e90}
\end{equation}
and so
\begin{equation}
E_{jk}(t) \cong 2mt e_{jk}(t) 
, 
\label{e100}
\end{equation}
where $e_{jk}$ is the degree--degree distribution for edges. Using the standard rate or master equation approaches \cite{Dorogovtsev:dms00,Krapivsky:krl00,Krapivsky:kr01} to this kind of networks (specifically, to the recursive networks growing due the preferential attachment mechanism), we write the following rate equations for $N_k(t)$ and $E_{jk}(t)$: 
\begin{eqnarray}
&&  \!\!\!\!\!\! \!\!\!\!\!\!
N_m(t+1) =  N_m(t) + 1 - m \frac{m+ma}{m(2+a)t}N_m(t) 
,
\nonumber
\\[5pt]
&&  \!\!\!\!\!\! \!\!\!\!\!\!
N_{k>m}(t+1) =  N_k(t) 
\nonumber
\\[5pt]
&& \!\!\!\!\!\! \!\!\!\!\!\!
+ m \frac{k-1+ma}{m(2+a)t}N_{k-1}(t) 
- m \frac{k+ma}{m(2+a)t}N_k(t)
\label{e110}
\end{eqnarray}
(note that the mean vertex degree in these networks is $\langle k \rangle = 2m$) and 
\begin{eqnarray}
&& E_{mk}(t+1) =  E_{mk}(t) + m \frac{k-1+ma}{m(2+a)t}N_{k-1}(t)
\nonumber
\\[5pt]
&& 
- m\left[\frac{m+ma}{m(2+a)t}E_{mk}(t) + \frac{k+ma}{m(2+a)t}E_{mk}(t) \right] 
,
\nonumber
\\[5pt]
&& 
E_{jk>m}(t+1) =  E_{jk}(t) 
\nonumber
\\[5pt]
&& 
+ m \left[\frac{k-1+ma}{m(2+a)t}E_{j,k-1}(t) 
        + \frac{j-1+ma}{m(2+a)t}E_{j-1,k}(t)\right]
\nonumber
\\[5pt]
&& 
- m \left[\frac{j+ma}{m(2+a)t}E_{jk}(t) 
        + \frac{k+ma}{m(2+a)t}E_{jk}(t)\right]
.
\label{e120}
\end{eqnarray}
We separately write out equations for the case of $k=m$.
Here we used the fact that the probability to attach a new vertex to a vertex $i$ of degree $k_i$ in this model is 
\begin{equation}
\frac{k_i+ma}{\sum_i (k_i+ma)} = \frac{k_i+a}{(\langle k \rangle+ma)t} 
= \frac{k_i+a}{m(2+a)t}
. 
\label{e125}
\end{equation}

Assuming a stationary regime at the limit $t\to\infty$ [$N_k(t) \cong tp_k$, $E_{jk}(t) \cong 2mt e_{jk}$], we reduce these equations 
to
\begin{eqnarray}
&& \!\!\!\! 
p_m =  1 - m \frac{m+ma}{m(2+a)}p_m 
, 
\nonumber
\\[5pt]
&& \!\!\!\! 
p_{k>m}
= m \frac{k-1+ma}{m(2+a)}p_{k-1} 
- m \frac{k+ma}{m(2+a)}p_k
\label{e130}
\end{eqnarray}
and
\begin{eqnarray}
&&  \!\!\!\!  \!\!\!\! 
2m(k+m+2+2ma+a)e_{mk} = (k-1+ma)p_{k-1} 
,
\nonumber
\\[5pt]
&&  \!\!\!\!  \!\!\!\! 
m(j+k+2+2ma+a)e_{jk>m}
\nonumber
\\[5pt]
&&  \!\!\!\!  \!\!\!\! 
= m(k-1+ma)e_{j,k-1} 
- m(j-1+ma)e_{j-1,k}
.
\label{e140}
\end{eqnarray}
To obtain the Pearson coefficient, see definition (\ref{e10}), we must find $\langle k^2 \rangle$, $\langle k^3 \rangle$, and $\langle jk \rangle_{\text{e}}$, where  
\begin{equation} 
\langle jk \rangle_{\text{e}} = \sum_{jk} jk e_{jk}
.
\label{e150}
\end{equation}
Multiplying both the sides of the second equation of the system (\ref{e130}) by  $k^2$ and $k^3$ and summing over $k$, and taking into account 
\begin{equation}
p_m = \frac{2 + a}{2 + a + m + ma}
,  
\label{e160}
\end{equation}
which follows from the first equation of the system (\ref{e130}), we obtain 
\begin{equation}
\langle k^2 \rangle = \sum_k k^2 p_k = \frac{m}{a}(2 + 5ma + a + 2m)
\label{e170}
\end{equation}
and 
\begin{eqnarray}
&&
\langle k^3 \rangle = \frac{m}{a(a-1)}
\nonumber
\\[5pt]
&&\!\!\!\!\!\!\!\!\!\!\!\!\!\!\!\!
\times[6(1{+}m) {+} a(5{+}21m{+}8m^2) {+} a^2(1{+}9m{+}16m^2)]
.
\label{e180}
\end{eqnarray}
respectively. 

To get $\langle jk \rangle_{\text{e}}$, we multiply both the sides of the second equation of the system (\ref{e140}) by $jk$ and sum them over $j$ and $k$. We also take into account the following general equality: 
\begin{equation}
\langle k \rangle_{\text{e}} = \sum_{jk} k e_{jk} = \frac{\langle k^2 \rangle}{\langle k \rangle}
, 
\label{e185}
\end{equation}
which gives  
\begin{equation}
\langle k \rangle_{\text{e}} = 
\frac{1}{2a}(2 + 5ma + a + 2m)
. 
\label{e190}
\end{equation}
As a result we find 
\begin{equation}
\langle jk \rangle_{\text{e}} = \frac{m}{a^2}[2+2m + a(5+7m) + a^2(2+7m)] 
,    
\label{e200}
\end{equation}
and so at $a>0$, i.e., $\gamma>3$, the numerator of definition (\ref{e10}) is 
\begin{eqnarray}
&&\!\!\!\!\!\!\!\!\!\!\!\!\!
\langle jk \rangle_{\text{e}} - \langle k \rangle_{\text{e}}^2 
\nonumber
\\[5pt]
&&\!\!\!\!\!\!\!\!\!\!\!\!\!
= 
\frac{m-1}{4a^2} [4(1+m) + 4a(1+2m) + a^2(1+3m)]  
.
\label{e205}
\end{eqnarray}
Importantly, this numerator is zero if $m=1$. So we see that $r_\infty(m=1)=0$ at least if $\gamma>3$. On the other hand, if $m>1$, the numerator is finite in the range $\gamma>3$. This shows that 
$r_\infty(m>1)$ must be zero 
at $3<\gamma\leq4$, since the denominator diverges if $\gamma\leq4$. Finally, at $\gamma\leq3$, $r_\infty=0$ at any $m$ [see the next section, Eq.~(\ref{e300})].  
Substituting relations (\ref{e170}), (\ref{e180}), and (\ref{e205}) into definitions (\ref{e10}) and (\ref{e20}) readily gives the resulting expression (\ref{e30}) for the stationary value $r_\infty$ of Pearson's coefficient.

\section{Size dependence} \label{zz}

\begin{figure}
\begin{center}
\scalebox{0.9}{\includegraphics[angle=0]{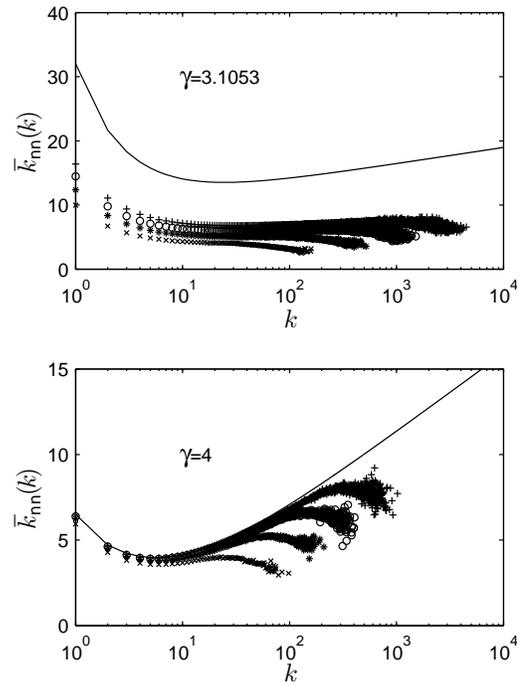}}
\end{center}
\caption{ 
Average degree of the nearest neighbours of a vertex versus the degree of this vertex for the scale-free recursive networks of various sizes with the degree distribution exponent values $3.1053$ and $4$, and $m=1$. The solid lines show the stationary dependences $\overline{k}_{\text{nn}}(k)$ in the infinite networks. These curves were find by numerical solving the stationary limits of the rate equations. Both these dependences, substituted into expressions (\ref{e210}) or (\ref{e220}) give exactly zero Pearson coefficient. 
The positive slope of $\overline{k}_{\text{nn}}(k)$ means assortative mixing, while the negative slope means disassortative mixing. The non-monotonous dependencies indicate the combination of these kind of degree--degree correlations, which, in this case, exactly compensate each other.  
The points show the results of simulation of the recursive networks of $10^3$ ($\times$), $10^4$ ($\times{}\hspace{-7.25pt}+$), $10^5$ ($\circ$), and $10^6$ ($+$) vertices. Note a very slow convergence of the results for finite networks to the infinite network limit at $\gamma$ close to $3$.   
} 
\label{f4}
\end{figure}

Equations~(\ref{e110}) and ~(\ref{e120}) allow one to derive the full size dependence of the Pearson coefficient. Instead of these cumbersome straightforward calculations, we obtain the asymptotic behavior of $r(t)$ in an easier way, using known results for the asymptotics of degree--degree correlations in these networks \cite{Krapivsky:kr01,Boguna:bpv03,Barrat:bp05,Dorogovtsev:dm01}. The derivation is based on the following expression of the Pearson coefficient in terms of the average vertex degree $\overline{k}_{\text{nn}}(k)$ of the nearest neighbors of the vertices of degree $k$:     
\begin{eqnarray}
&& 
r = 
\frac{\langle k \rangle\sum_k k(k-1) p_k 
[\overline{k}_{\text{nn}}(k) - \langle k^2 \rangle/\langle k \rangle]}{\langle k \rangle\langle k^3 \rangle - \langle k^2 \rangle^2} 
\label{e210}
\\[5pt]
&& 
=
\frac{\langle k \rangle \sum_k k^2 p_k  
\overline{k}_{\text{nn}}(k) - \langle k^2 \rangle^2}{\langle k \rangle\langle k^3 \rangle - \langle k^2 \rangle^2} 
.
\label{e220}
\end{eqnarray}
This expression directly follows from definition (\ref{e10}). The leading asymptotics of $\overline{k}_{\text{nn}}(k)$ can be obtained by using the known exact asymptotics of $n_{kl}$ for these recursive networks \cite{Krapivsky:kr01}. Here $n_{kl}$ is the probability that a descendant vertex of degree $k$ is connected to an ascendant vertex of degree $l$. This quantity satisfies the following relations: 
\begin{eqnarray}
&&
\sum_l n_{kl} = p_k 
,
\nonumber
\\[5pt]
&&
\sum_l (n_{kl} + n_{lk}) = kp_k 
, 
\nonumber
\\[5pt]
&&
\sum_l l(n_{kl} + n_{lk}) = kp_k \overline{k}_{\text{nn}}(k)
.
\label{e230}
\end{eqnarray}
At $m=1$, according to Ref.~\cite{Krapivsky:kr01}, 
\begin{eqnarray}
&&
n_{kl} \sim l^{\gamma-2}k^{1-2\gamma}  \ \ \text{if} \ \  1 \ll l \ll k 
, 
\nonumber
\\[5pt]
&&
n_{kl} \sim l^{1-\gamma}k^{-2}  \ \ \ \ \text{if} \ \  1 \ll k \ll l
.
\label{e240}
\end{eqnarray}
For $\gamma>3$ and any $m$, $\overline{k}_{\text{nn}}(k,t)$ approaches a stationary limit as the network grows. Figure~\ref{f4} demonstrates this relaxation for two values of the degree distribution exponent $\gamma$, namely, for $\gamma=3.1053$ and $\gamma=4$ in the case of $m=1$. We can approximate the deviation from the infinite network limit by   
\begin{equation}
[\overline{k}_{\text{nn}}(k) - \overline{k}_{\text{nn}}(k,t)]kp_k \sim \int_{k_{\text{cut}}}^\infty dl\, l (n_{kl} + n_{lk})
, 
\label{e250}
\end{equation}
where $k_{\text{cut}} = k_{\text{cut}}(t)$ is the time dependent cutoff of the degree distribution. In these networks, $k_{\text{cut}}(t) \sim t^{1/(\gamma-1)}$. Substituting Eq.~(\ref{e240}) into Eq.~(\ref{e250}) results in 
\begin{eqnarray}
&&
\overline{k}_{\text{nn}}(k) - \overline{k}_{\text{nn}}(k,t) 
\nonumber
\\[5pt]
&&
\sim c_1 k^{\gamma-3} t^{(3-\gamma)/(\gamma-1)} + c_ 2 k^{2\gamma-3} t^{-(2\gamma-3)/(\gamma-1)}
\label{e260}
\end{eqnarray}
at large degrees $k<k_{\text{cut}}$ for the networks with $\gamma>3$. Here $c_1$ and $c_2$ are constants depending on $\gamma$, which we do not calculate. 

Similarly, for $2<\gamma<3$, where $\overline{k}_{\text{nn}}(k,t)$ diverges as $t \to \infty$ at any $m$, we can write an asymptotic estimate 
\begin{equation}
\overline{k}_{\text{nn}}(k,t)kp_k \sim \int_m^{k_{\text{cut}}} dl\, l (n_{kl} + n_{lk})
.  
\label{e270}
\end{equation}
Substituting Eq.~(\ref{e240}) into Eq.~(\ref{e270}) we obtain  
\begin{equation}
\overline{k}_{\text{nn}}(k,t) \sim c_3 k^{\gamma-3} t^{(3-\gamma)/(\gamma-1)} + c_4 k^{2\gamma-3} t^{(3-2\gamma)/(\gamma-1)}
\label{e280}
\end{equation}
at large degrees $k<k_{\text{cut}}$ for the networks with $2<\gamma<3$. Here $c_3$ and $c_4$ are some constants 
\cite{bookmark2}. 

We substitute asymptotics (\ref{e260}) and (\ref{e280}) into expression (\ref{e220}) for the Pearson coefficient 
and take into account the leading terms in the numerator and the denominator. The combinations of these leading terms are different in different areas of $\gamma$ and $m$, since the quantities $\overline{k}_{\text{nn}}(k,t)$, $\langle k^2 \rangle$, and $\langle k^3 \rangle$ in expression~(\ref{e220}) change their asymptotic behavior at two special points, namely, $\gamma=3$ and $\gamma=4$. In particular, at these points ($\gamma=3$ and $4$) the second and, respectively, the third moments of the degree distribution become divergent. 
For example, if $2<\gamma<3$ at any $m$, relation~(\ref{e220}) with substituted asymptotics (\ref{e260}) takes the form: 
\begin{eqnarray}
&&\!\!\!\!\!
r(t) \sim 
\left[\left(c'_1 t^{(3-\gamma)/(\gamma-1)} \int^{t^{1/(\gamma-1)}} \!\!\!\!\!\!\!\!dk\, k^2 k^{-3+\gamma}k^{-\gamma} \right.\right.
\nonumber
\\[5pt]
&&\!\!\!\!\! 
+\! \left.c'_2 t^{(3-2\gamma)/(\gamma-1)}\!\!\int^{t^{1/(\gamma-1)}} \!\!\!\!\!\!\!\!\!\!\!\!\!\!\!\!\!\!dk\,k^2 k^{2\gamma-3} k^{-\gamma}\!\right)  
\!-\! c'_3\!\!\left.\left(\! \int^{t^{1/(\gamma-1)}} \!\!\!\!\!\!\!\!\!\!\!\!\!\!\!\!dk\,k^{2-\gamma}\right)^{\!\!2}\right]
\nonumber
\\[5pt]
&&\!\!\!\!\!
\left/\left(2m\int^{t^{1/(\gamma-1)}} \!\!\!\!\!\!dk\,k^{3-\gamma}\right)\right.
, 
\label{e290}
\end{eqnarray}
where $c'_1$, $c'_2$ and $c'_3$ are constants. 
This leads to the result: 
\begin{eqnarray}
&&\!\!\!\!\!\!\!
r(t)  
\nonumber
\\[5pt]
&&\!\!\!\!\!\!
\sim \!\frac{[c''_1 \ln t\,\,t^{(3-\gamma)/(\gamma-1)} + c''_2 t^{(3-\gamma)/(\gamma-1)}] - c''_3 t^{2(3-\gamma)/(\gamma-1)}}{t^{(4-\gamma)/(\gamma-1)}} 
\nonumber
\\[5pt]
&&\!\!\!\!\!\!
\sim - 
t^{-(\gamma-2)/(\gamma-1)}
, 
\label{e300}
\end{eqnarray}
where $c''_1$, $c''_2$ and $c''_3$ are constants.
In a similar way, we derive the other asymptotics listed in Eqs.~(\ref{e50}) and (\ref{e60}). 

Interestingly, both the terms in Eqs.~(\ref{e260}) and (\ref{e280}) give contributions of the same order of magnitude to $r(t)$, see the numerator of Eq.~(\ref{e300}). 
Note 
that we suppose that for $m>1$ the form of the asymptotics of $\overline{k}_{\text{nn}}(k,t)$ is the same as in Eqs.~(\ref{e260}) and (\ref{e280}). 
To verify our assumption, we inspected the corresponding results for $\overline{k}_{\text{nn}}(k,t)$ in Ref.~\cite{Barrat:bp05} and found that the asymptotic behavior should be similar at different $m$ (apart of numerical coefficients) if exponent $\gamma$ is fixed. In addition, we checked that results~(\ref{e50}) also can be derived by using the asymptotics of $\overline{k}_{\text{nn}}(k,t)$ from Ref.~\cite{Barrat:bp05}. This confirms our conclusions about the asymptotics of $r(t)$.


\section{Discussion and conclusions}
\label{conclusions}

Our result, namely zero Pearson coefficient of random recursive trees 
at any $\gamma>2$, naturally leads to the following questions. What is the class of trees that have zero Pearson coefficient? Is Pearson's coefficient zero for any infinite recursive tree? At present, we cannot answer the first question. As for the second question, the answer is negative. Indeed, as a counter-example, we present the simplest infinite recursive tree with a non-zero Pearson coefficient. This is a star, which is a tree of $t>2$ vertices, including $t-1$ leafs and the hub of degree $t-1$. For star of any size, clearly, $r=-1$.  


Thus we have studied the Pearson coefficient in strongly correlated growing networks. 
They form a representative class of networks with strong structural correlations including pairwise correlations between degrees of the nearest-neighbor vertices. Despite these correlations, we have found that in a wide range of infinite correlated networks the Pearson coefficient approaches zero. For any infinite random recursive tree whose growth is driven by arbitrary linear preferential attachment of new vertices, we observed zero Pearson coefficient. These networks include random recursive trees with rapidly decaying and even exponential degree distribution, where the third moment of the degree distribution is finite. 
So here zero value of Pearson's coefficient demands zero correlation function in the numerator of definition ({\ref{e10}}). 
This 
surprising equality to zero indicates an exact mutual compensation of assortative and disassortative degree--degree correlations.  
In this respect, the recursive trees is a very special case of random recursive networks \cite{bookmark4}.  

We have investigated the size dependence of Pearson's coefficient in the growing networks. 
We have found that the size effect is significant even for very large networks. We have shown that a growing network during its evolution may demonstrate essential and even non-monotonous variation of Pearson's coefficient. 
Due to this marked size dependence, it is hardly feasible 
to use this integral characteristic of correlations for quantitative comparison of different real-world networks. 
Instead, for this purpose, one has to use more informative characteristics, for example $\overline{k}_{\text{nn}}(k)$.

\begin{acknowledgments} 

This work was partially supported by the following projects PTDC: FIS/71551/2006, FIS/108476/2008, and SAU-NEU/103904/2008, and also by SOCIALNETS EU project. The authors thank M.~Barroso for his help with 
administration of 
computing facilities used for numerical simulations. 

\end{acknowledgments}




\begin{thebibliography}{99}
\bibitem{bookmark1} 
Note that the equality $\langle k^n \rangle_{\text{e}}=\langle k^{n+1} \rangle/\langle k \rangle$ is valid for an arbitrary network.
\bibitem{bookmark3} 
Note that the numerator in definition (\ref{e10}) also diverges if $\gamma\leq3$. This divergence, however, is slower than that of the denominator, see Eq.~(\ref{e300}) which leads to zero Pearson's coefficient in this range.
\bibitem{bookmark2} 
Note that there is some difference between asymptotics 
(\ref{e280}) and the corresponding result in Ref.~\cite{Barrat:bp05}. The difference is apparently due to the fact that we derived these expression by using the exact asymptotics taken from Ref.~\cite{Krapivsky:kr01} while the result of Ref.~\cite{Barrat:bp05} was obtained by using the continuous approximation for the rate equations. Nonetheless, one can easily check that for $2<\gamma<3$, the expression for $\overline{k}_{\text{nn}}(k,t)$ from Ref.~\cite{Barrat:bp05} leads to the same asymptotic behavior of the Pearson coefficient as in the present paper.
\end{thebibliography}
\end{document}